\documentclass[journal,comsoc]{IEEEtran}
\usepackage{cite}
\usepackage{graphicx}
\usepackage{amsmath}
\usepackage{amsthm}
\usepackage{times}
\usepackage{graphicx}
\usepackage{bm}
\usepackage{amssymb}
\usepackage[center]{caption2}
\usepackage{stfloats}
\usepackage{cases}
\usepackage{subfigure}
\usepackage{array}
\usepackage{setspace}
\usepackage{fancyhdr}
\usepackage{colortbl}
\usepackage{CJK}
\usepackage{tikz}
\usepackage{float}
\usepackage{color}
\usepackage{bm}
\usepackage{mathrsfs}
\usepackage{diagbox}
\usepackage{cuted}
\usepackage{url}

\newtheorem{theorem}{Theorem}

\newtheorem{remark}{Remark}

\def\proof{\noindent\hspace{2em}{\itshape Proof: }}
\def\endproof{\hspace*{\fill}~$\square$\par\endtrivlist\unskip}

\allowdisplaybreaks[4]

\newcaptionstyle{mystyle2}{%
\captionlabel $.$ \, \captiontext \par} \captionstyle{mystyle2}

\begin{document}
\title{Performance Analysis of Intelligent Reflecting Surface Aided Communication Systems}
\author{\IEEEauthorblockN{Qin Tao, Junwei Wang, and Caijun Zhong}
\thanks{Q. Tao, J. Wang, and C. Zhong are with the College of Information Science and Electronic Engineering, Zhejiang University, Hangzhou, China (email: caijunzhong@zju.edu.cn).}
}

\maketitle

\begin{abstract}
This letter presents a detailed performance analysis of the intelligent reflecting surface (IRS) aided single-input single-output communication systems, taking into account of the direct link between the transmitter and receiver. A closed-form upper bound is derived for the ergodic capacity, and an accurate approximation is obtained for the outage probability. In addition, simplified expressions are presented in the asymptotic regime. Numerical results are provided to validate the correctness of the theoretical analysis. It is found that increasing the number of reflecting elements can significantly boost the ergodic capacity and outage probability performance, and a strong line-of-sight component is also beneficial. In addition, it is desirable to deploy the IRS close to the transmitter or receiver, rather than in the middle.
\end{abstract}


\section{Introduction}
The intelligent reflecting surface (IRS), which can manipulate the {propagation} channel into a favorable shape, has been regarded as a promising technology for the next generation wireless communication systems. As such, it has received considerable interests from both the industry and academia {\cite{chuang1,chuang2}}.

Thus far, most of the works on IRS focus on the design of the phase shift matrix and the transmit beamformer \cite{Q.Wu1,Q.Wu2,Huang,jinshi,zhou,shuowen,X.Hu}. For multiple-input single-output (MISO) systems, the joint active and passive beamfomer design problem was studied in \cite{Q.Wu1}, and the impact of discrete phase shift was further considered in \cite{Q.Wu2}. Meanwhile, the energy efficiency of the systems was characterized in \cite{Huang}. Later on, a statistical channel state information (CSI) based design framework was proposed in \cite{jinshi}. The more general multiple-input multiple-output (MIMO) system was considered in \cite{shuowen}, while the multi-cast MISO scenario was addressed in \cite{X.Hu,zhou}.

 {While the aforementioned works have improved our knowledge of IRS-aided communication systems, few works have studied the analytical performance of IRS aided systems}. In \cite{cui1,cui2}, the outage probability and ergodic spectral efficiency of the MISO systems was studied, assuming that the channel between the transmitter and IRS is deterministic.  {In \cite{SISO,general}, the outage probability and achievable rate of single-input single-output (SISO) system was considered, assuming that the direct link between the transmitter and receiver does not exist.}

In practice, the IRS is usually deployed in a position with line-of-sight (LOS) to both transmitter and receiver, it is desirable to adopt the Rician fading model. Motivated by this, in this letter, we present a detailed performance analysis of IRS-aided SISO systems in Rician fading channels, taking into account of the direct channel between the transmitter and receiver. Closed-form expressions are derived for the ergodic capacity upper bound and outage probability approximation of the system. In addition, simplified expressions are obtained in the asymptotic regime. The findings of the letter suggest that the IRS can provide an effective signal-to-noise ratio (SNR) gain of $N^2$, where $N$ is the number of reflecting elements, and a diversity order of $N+1$ can be achieved.


\section{System Model}
\begin{figure}[tbp]
    \centering
    \includegraphics[width=0.5\linewidth]{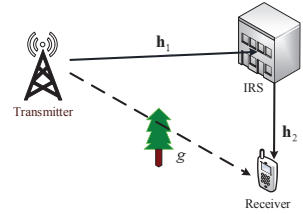}
	\vspace{-8mm}
    \caption{IRS-aided communication systems.}
    \label{IRS}
\end{figure}
We consider a three-node communication system consisting of a single antenna transmitter T, a single antenna receiver R and an IRS with $N$ reflecting elements, as illustrated in Fig.~\ref{IRS}.

Since the receiver can overhear the signal from both the T-IRS-R and T-R links, the received signal $y$ at the R can be expressed as
\begin{equation}
    y = \sqrt{P}\left({\mathbf{h}_2^T \boldsymbol{\Phi} \mathbf{h}_1}+{g}\right)x + n,
\end{equation}
where $x$ denotes the normalized transmit signal with unit energy, $g$ denotes the T-R channel, the $N\times 1$ vectors $\mathbf{h}_1$ and $\mathbf{h}_2$ denote the T-IRS and IRS-R channels, respectively. Also, the $N\times N$ matrix $\mathbf{\Phi}\triangleq {\sf diag}\{\boldsymbol{\theta}\}$ denotes the phase shift matrix, where $\boldsymbol{\theta}=[e^{j\theta_1}, e^{j\theta_2}, \dots,e^{j\theta_N}]$ with $\theta_{n}\in [0,2\pi)$ representing the phase shift of the $n$-th IRS reflecting element. Finally, $n$ represents the additive white Gaussian noise with zero mean and variance $N_0$.

Since the IRS is deployed at a position that has LOS to both T and R, $\mathbf{h}_1$ and $\mathbf{h}_2$ are modeled by Rician fading. In contrast, there is no LoS path between T and R, hence, $g$ is modeled by Rayleigh fading. Therefore, the channel coefficients can be expressed as
\begin{align}
    &\mathbf{h}_l = \frac{1}{\sqrt{d_l^{\alpha_l}}}\left(\sqrt{\frac{K_l}{K_l+1}} \mathbf{\Bar{h}}_l + \sqrt{\frac{1}{K_l+1}} \mathbf{\Tilde{h}}_l\right), \quad l=1,2,\notag\\
    &g = \frac{1}{\sqrt{d_3^{\alpha_3}}} {\Tilde{h}}_3,\notag
\end{align}
where $d_l$ and ${\alpha_l}$, $l\in\{1,2,3\}$, denote the distance and path loss exponent of the corresponding channel, while $K_i$, $i\in\{1,2\}$ denotes the Rician factor. Also, $\mathbf{\Bar{h}}_l$, $l\in\{1,2\}$, denotes the normalized LoS component, and $\mathbf{\Tilde{h}}_l$, $l\in\{1,2,3\}$ denotes the normalized non-LOS component.


Assuming perfect CSI at the IRS, the optimal phase shift matrix $\mathbf{\Phi}$ maximizing the received SNR is give by \cite{J.Gao}
\begin{align}
\theta_n = \angle\frac{g}{{h}_{2,n}{h}_{1,n}},
\end{align}
where $\angle x$ denotes the phase of complex number $x$, $h_{1,n}$ and ${h}_{2,n}$ are the $n$-th element of vector ${\bf h}_1$ and ${\bf h}_2$, respectively. Then, the maximum SNR of the system can be obtained as
\begin{align}\label{gammax}
\gamma_{\sf max} = \gamma_0\left( \sum_{n=1}^N |{h}_{2,n}| |{h}_{1,n}| + |g| \right)^2,
\end{align}
where $\gamma_0=P/N_0$ represents the transmit SNR.
%
%

\section{Performance Analysis}
In this section, we provide a detailed analysis of the achievable systems performance. Specifically, two important metrics, i.e., ergodic capacity and outage probability, are considered. We start with the ergodic capacity.

\subsection{Ergodic Capacity}
The ergodic capacity of the system can be expressed as
\begin{align}
    C = \mathbb{E} \left\{ \log_2 \left( 1+\gamma_{\sf max} \right) \right\}.
\end{align}
Since the exact distribution of $\gamma_{\sf max} $ is intractable, it is challenging to characterize the exact ergodic capacity. Therefore, we resort to a tractable bound with the help of Jensen's inequality, and have the following key results:
\begin{theorem}
The ergodic capacity of the system is upper bounded by
 $C_{\sf up}$ given in Eq. (\ref{the:1}), shown on the top of the next page,
\begin{figure*}
\begin{multline}\label{the:1}
C_{\sf up}=\log_2\Bigg\{1+\gamma_0\bigg[\frac{1}{d_3^{\alpha_3}}+\frac{N}{d_1^{\alpha_1}d_2^{\alpha_2}}+\\
\frac{N(N-1)\pi^2}{16d_1^{\alpha_1}d_2^{\alpha_2}(K_1+1)(K_2+1)} \left( \text{L}_{\frac{1}{2}}(-K_1) \text{L}_{\frac{1}{2}}(-K_2) \right)^2+\sqrt{\frac{N^2\pi^3}{8d_1^{\alpha_1}d_2^{\alpha_2}d_3^{\alpha_3}(K_1+1)(K_2+1)}}\text{L}_{\frac{1}{2}}(-K_1)\text{L}_{\frac{1}{2}}(-K_2)\bigg]\Bigg\}
\end{multline}
\hrule
\end{figure*}
where $\text{L}_{\frac{1}{2}}(\cdot)$ denotes the Laguerre polynomial \cite{Handbook}.
\end{theorem}
\proof See Appendix A. \endproof

Theorem 1 provides a closed-form expression involving only elementary functions, which is applicable for arbitrary system configurations, thereby enabling efficient evaluation of the ergodic capacity performance. In addition, it also facilitates the characterization of the impact of key parameters on the ergodic capacity. In particular, we have the following observations:
\begin{remark}
The ergodic capacity upper bound increases monotonically with $N$. For sufficiently large $N$, the ergodic capacity $C_{\sf up}$ is mainly dominated by the term $\log_2\gamma_1N^2$, where $\gamma_1=\gamma_0\frac{\left( \text{L}_{\frac{1}{2}}(-K_1) \text{L}_{\frac{1}{2}}(-K_2) \pi\right)^2}{16d_1^{\alpha_1}d_2^{\alpha_2}(K_1+1)(K_2+1)}$, indicating an effective SNR gain of $N^2$ . This can be explained by the fact that IRS not only achieves the beamforming gain, but also attains the inherent aperture gain by collecting more signal power.
\end{remark}

\begin{remark}
Observing that $\text{L}_{\frac{1}{2}}(-K)$ is a monotonically increasing function of $K$, which suggests that a strong LOS component would enhance the ergodic capacity. In addition, the ergodic capacity upper bound $C_{\sf up}$ is a symmetric function with respect to the $K_1$ and $K_2$, indicating the identical impact of the two hop channels.
\end{remark}

\subsection{Outage Probability}
In this subsection, we analyze the outage probability of the system, which is defined as the probability of the instantaneous SNR  {that} falls below a pre-defined threshold $\gamma_{\sf th}$. Mathematically, it is given by
\begin{align}
P_{\sf out} = {\sf Prob}\left(\gamma_{\sf max}\leq \gamma_{\sf th}\right).
\end{align}

Since the exact distribution of $\gamma_{\sf max}$ is unknown, we resort to tight approximations of the outage probability, and we have the following important result:
\begin{theorem}
 {When $N\rightarrow \infty$}, the outage probability can be approximated as Eq. (\ref{the:2}) shown on the top of the next page, where
\begin{align}
\mu = \frac{\pi}{4\sqrt{d_1^{\alpha_1}d_2^{\alpha_2}(K_1+1)(K_2+1)}} \text{L}_{\frac{1}{2}}(-K_1)\text{L}_{\frac{1}{2}}(-K_2),\notag
\end{align}
and
\begin{align}
\sigma^2 = \frac{1}{d_1^{\alpha_1}d_2^{\alpha_2}}\left[1 - \frac{\pi^2\left( \text{L}_{\frac{1}{2}}(-K_1)\text{L}_{\frac{1}{2}}(-K_2) \right)^2}{16(K_1+1)(K_2+1)}  \right].
\end{align}
\begin{figure*}
\begin{align}\label{the:2}
P_{\sf out,1} \approx  \frac{1}{2}+\frac{1}{2}\text{erf}\left(\frac{\sqrt{\frac{\gamma_{\sf th}}{\gamma_0}}-N\mu}{\sqrt{2N\sigma^2}}\right) - \frac{1}{2\sqrt{1+N\sigma^2{d_3^{\alpha_3}}}}\exp\left(\frac{-(\sqrt{\frac{\gamma_{\sf th}}{\gamma_0}}-N\mu)^2}{2({1}/{d_3^{\alpha_3}}+N\sigma^2)}\right)\left[ 1+\text{erf}\left(\frac{\sqrt{\frac{\gamma_{\sf th}}{\gamma_0}}-N\mu}{\sqrt{2N\sigma^2(1+N\sigma^2{d_3^{\alpha_3}})}}\right) \right].
\end{align}
\hrule
\end{figure*}

\end{theorem}
\proof See Appendix B.\endproof
Theorem 2 presents a closed-from approximation for the outage probability, which consists of only elementary functions, hence can be efficiently evaluated. Also, we observe that both $\mu$ and $\sigma^2$ are symmetric functions with respect to $K_1$ and $K_2$, which implies $K_1$ and $K_2$ have identical impact on the outage probability. In addition, although Theorem 2 is obtained with the assumption of large $N$, as will be shown through numerical results, the approximation turns out to be sufficiently tight even for moderate $N$.

To gain further insight, we now look into the high SNR regime, and we have the following important result:
\begin{theorem}\label{theohigh}
 {When $\gamma_0 \rightarrow \infty$}, the outage probability can be accurately approximated by
\begin{align}
P_{\sf out,2} \approx
\left[\frac{\sqrt{\pi}a^{N}d_1^{\alpha_1N}d_2^{\alpha_2N}d_3^{\alpha_3}}{\Gamma\left(N+\frac{3}{2}\right)\left( N+1 \right)!}\right]\bigg(\frac{2\gamma_0}{\gamma_{\sf th}} \bigg)^{-(N+1)},
\end{align}
where
\begin{align}
&a = (K_1+1)(K_2+1) e^{-(K_1+K_2)}\bigg[ {\sf Ei}(K_1)+{\sf Ei}(K_2)-2\gamma\notag\\
&-\ln K_1-\ln K_2+2K_{0}\left(2\sqrt{\frac{(K_1+1)(K_2+1)}{d_1^{\alpha_1}d_2^{\alpha_2}\gamma_0}}\right)\bigg],\notag
\end{align}
$\gamma$ denotes the Euler's constant, ${\sf Ei}(x)$ is the exponential integral\cite{NIST}, and $K_n(x)$ is the modified Bessel function of the second kind\cite{NIST}.
\end{theorem}
\proof
See Appendix \ref{outhigh}.
\endproof
Theorem \ref{theohigh} suggests that a diversity order of $N+1$ is achieved. Moreover, the impact of LOS component is mainly reflected on the achievable coding gain of the system.

\section{Simulation Results}
In this section, numerical results are presented to validate the theoretical analysis in the previous section. Unless otherwise specified, the following set of parameters are used in the simulations.  {As in \cite{Huang}, we focus on the sub-6G scenario, the system has a bandwidth of 180KHz, and the noise power spectrum density is -173 dBm/Hz.} The distances are set to be $d_1=150\ \mathrm{m}$, $d_2=150\ \mathrm{m}$ and $d_3=200\ \mathrm{m}$, with path loss exponents given by $\alpha_0=3.5$, $\alpha_1=\alpha_2=2.0$, respectively.
The reference path loss at the reference distance $d_0\!=\!1$ m is set to be $-30$ dB and the outage threshold $\gamma_{\sf th}=10$ dB. Also, we set $K_1 = K_2 = K$. 

\begin{figure}[tbp]
	\centering
	\includegraphics[width=0.8\linewidth]{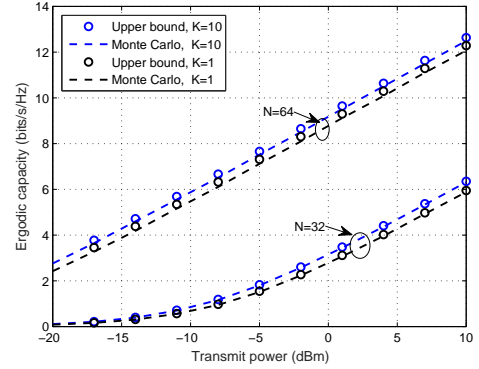}
	\caption{Impact of $K$ and $N$ on the ergodic capacity.}
	\label{C_vs_K}
\end{figure}

Fig. \ref{C_vs_K} illustrates the ergodic capacity of the system with different $N$. As can be readily observed, the ergodic capacity upper bound is tight for all configurations, indicating the accuracy of the closed-form expression in Theorem 1. In addition, we observe the intuitive result that the ergodic capacity increases monotonically with the number of IRS reflecting elements $N$ and Rician factor $K$. 

\begin{figure}[tbp]
	\centering
	\includegraphics[width=0.8\linewidth]{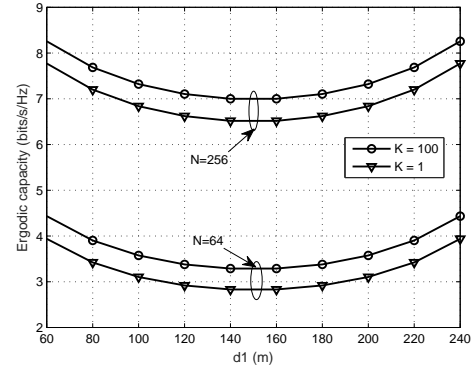}
	\caption{Impact of IRS location on the ergodic capacity.}
	\label{C_vs_d}
\end{figure}

Fig. \ref{C_vs_d} shows the impact of IRS location on the ergodic capacity assuming the IRS is deployed between the line segment of T and R, i.e., $d_1+d_2=300$ m. As can be readily observed, the ergodic capacity is a symmetric function with respect to $d_1$ and $d_2$. The minimum is achieved when the IRS is deployed in the middle of the T and R. The above result indicates that it better to deploy the IRS in the vicinity of either T or R to obtain higher capacity.

\begin{figure}[tbp]
	\centering
	\includegraphics[width=0.8\linewidth]{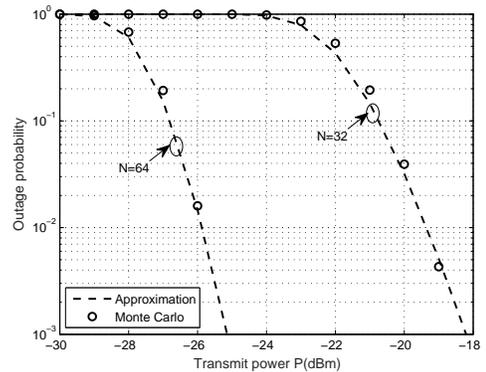}
	\caption{Outage approximation $P_{\sf out,1}$ with $K=1$. }
	\label{eq:outage1}
\end{figure}

\begin{figure}[tbp]
	\centering
	\includegraphics[width=0.8\linewidth]{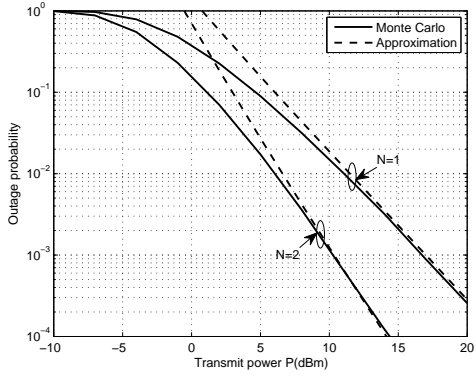}
	\caption{High SNR outage approximation $P_{\sf out,2}$ with $K=1$.}
	\label{eq:outage2}
\end{figure}

Fig. \ref{eq:outage1} plots the outage probability of the system with $K=1$. It can be observed that the approximation works well, even for moderate value of $N$. In addition, we see that the outage probability decreases sharply when the SNR increases, and the decreasing rate is faster with a larger $N$.

Fig. \ref{eq:outage2} validates the accuracy of $P_{\sf out,2}$ in the high SNR regime when $K =1$. As readily can be seen, the high SNR approximation is quite accurate, thus validates the correctness of $P_{\sf out,2}$. Also, a diversity order of $N+1$ is observed, indicating the great benefit of increasing $N$ in terms of outage performance.

\section{conclusion}
This paper has studied the ergodic capacity and outage probability of IRS-aided SISO systems under mixed Rayleigh and Rician fading channels. Closed-from expressions are derived for the ergodic capacity upper bound and outage approximation, which provide efficient means to evaluate the system performance. Moreover, concise expressions are obtained in the asymptotic regime, which sheds lights on the impact of key parameters on the system performance. It is revealed that the use of IRS contributes to an effective SNR gain of $N^2$, and the diversity order can be increased to $N+1$. Also, the position of IRS also has significant impact on the system performance, it is desirable to deploy the IRS close to the transmitter or receiver, and with strong LOS path.

\begin{appendices}
\section{Proof of Theorem 1}

Applying the Jensen's inequality, we have the upper bound of the capacity as
\begin{align}\label{C_approx}
 C\leq C_{\sf up} = \log_2 \left( 1+ \mathbb{E} \left\{ \gamma_{\sf max} \right\} \right).
\end{align}
The remaining task is to compute $\mathbb{E} \left\{ \gamma_{\sf max}\right\} $. Using the relationship in (\ref{gammax}) and applying the binomial expansion theorem, we have
\begin{align}
     \mathbb{E} \left\{ \gamma_{\sf max}\right\} = &\underbrace{ \mathbb{E} \left\{|g|^2 \right\} }_{x_1}+\underbrace{ \mathbb{E} \left\{ \left(\sum_{n=1}^N|{h}_{2,n}||{h}_{1,n}|\right)^2 \right\} }_{x_2}+\notag\\
     &\underbrace{ 2\mathbb{E} \left\{ \sum_{n=1}^N|{h}_{2,n}||{h}_{1,n}||g| \right\} }_{x_3}.
\end{align}
We now calculate $x_1$, $x_2$ and $x_3$ one-by-one.

1) Computing $x_1$: Obviously, we have $x_1 = \frac{1}{d_3^{\alpha_3}}$.

2) Computing $x_2$: Using the binomial expansion, $x_2$ can be further expanded as
\begin{align}
    x_2 =& \mathbb{E}\left\{\sum_{n=1}^N |{h}_{2,n}|^2 |{h}_{1,n}|^2\right\} + \notag\\
    &\mathbb{E}\left\{\sum_{n=1}^N \sum_{\substack{j=1 \\ j\ne n}}^N |{h}_{2,n}| |{h}_{1,n}| |h_{2,j}| |h_{1,j}|\right\}.
\end{align}
It is easy to see that $\mathbb{E}\left\{\sum_{n=1}^N |{h}_{2,n}|^2 |{h}_{1,n}|^2\right\} = \frac{N}{d_1^{\alpha_1}d_2^{\alpha_2}}$. Regarding $\mathbb{E}\left\{\sum_{n=1}^N \sum_{\substack{j=1 \\ j\ne n}}^N |{h}_{2,n}| |{h}_{1,n}| |h_{2,j}| |h_{1,j}|\right\}$, noticing that, for the Rician variable $|h_{l,n}|$, $l\in\{0,1\}$,
\begin{align}
\mathbb{E}\{|h_{l,n}|\} =\sqrt{\frac{\pi}{4d_l^{\alpha_l}(K_l+1)}} \text{L}_{\frac{1}{2}}(-K_l),
\end{align}
and observing the fact that $\mathbf{h}_1$ and $\mathbf{h}_2$ are independent, we obtain
\begin{align}
&\mathbb{E}\left\{\sum_{n=1}^N \sum_{\substack{j=1 \\ j\ne n}}^N |{h}_{2,n}| |{h}_{1,n}| |h_{2,j}| |h_{1,j}|\right\}\notag\\
= &\frac{N(N-1)\pi^2}{16d_1^{\alpha_1}d_2^{\alpha_2}(K_1+1)(K_2+1)} \left( \text{L}_{\frac{1}{2}}(-K_1) \right)^2 \left( \text{L}_{\frac{1}{2}}(-K_2) \right)^2.
\end{align}

3) Computing $x_3$: For the Rayleigh variable $|g|$, we have
\begin{align}
\mathbb{E}\{|g|\} = \sqrt{\frac{\pi}{2d_3^{\alpha_3}}}.
\end{align}
As such, $x_3$ can be calculated as
\begin{align}
x_3 = \sqrt{\frac{\pi^3N^2}{8d_1^{\alpha_1}d_2^{\alpha_2}d_3^{\alpha_3}(K_1+1)(K_2+1)}}\text{L}_{\frac{1}{2}}(-K_1)\text{L}_{\frac{1}{2}}(-K_2).
\end{align}

To this end, pulling $x_1$, $x_2$ and $x_3$ together yields the desired result.

\section{Proof of Theorem 2}
According to the definition, the outage probability can be transformed into
\begin{align}
P_{\sf out}(\gamma_{\sf th}) = {\sf Prob}\left(z \leq \sqrt{\frac{\gamma_{\sf th}}{\gamma_0}}\right),
\end{align}
where $z\triangleq u + |g|$ with $u\triangleq \sum_{n=1}^N |{h}_{2,n}| |{h}_{1,n}|$. For sufficiently large $N$, $u$ can be accurately approximated by the normal distribution according to the central  {limit} theorem, i.e., $u \sim \mathcal{N}(N\mu, N\sigma^2)$, where
\begin{align}
\mu &= \mathbb{E}\{|h_{2,n}||h_{1,n}|\}\\
 &= \frac{\pi}{4\sqrt{d_1^{\alpha_1}d_2^{\alpha_2}(K_1+1)(K_2+1)}} \text{L}_{\frac{1}{2}}(-K_1)\text{L}_{\frac{1}{2}}(-K_2),\notag
\end{align}
and
\begin{align}
    &\sigma^2 = \mathbb{D}\{|h_{2,n}||h_{1,n}|\}= \frac{1}{d_1^{\alpha_1}d_2^{\alpha_2}}\\
    &\left[1 - \frac{\pi^2}{16(K_1+1)(K_2+1)}  \left( \text{L}_{\frac{1}{2}}(-K_1)\text{L}_{\frac{1}{2}}(-K_2) \right)^2\right].\notag
\end{align}
Since $z$ is the sum of independent random variable $u$ and $|g|$, its cumulative distribution function (CDF) can be calculated via
\begin{align}
    F_z(z) &\approx \int_{-\infty}^{\infty} f_u(z-x) F_{|g|}(x)\text{d}x,
\end{align}
where $f_u(x)=\frac{1}{\sqrt{2\pi N \sigma^2}} \exp\left[\frac{-(x-N\mu)^2}{2N\sigma^2}\right]$ is the probability density function (PDF) of $u$ and
$F_{|g|}(x)=1 - \exp\left(\frac{-d_3^{\alpha_3}x^2}{2}\right)$ denotes the CDF of Rayleigh variable $|g|$.

To this end, the desired result can be obtained after some algebraic manipulations.

\section{Proof of Theorem \ref{theohigh}}\label{outhigh}
Recall $\gamma_{\sf max}$ in Eq. (2), it can be interpreted as the effective SNR of an equal gain combining SIMO system. Then, according to \cite{diversity}, in order to obtain the outage approximation in the high SNR regime, it is sufficient to characterize the behavior of the probability density function (PDF) of the SNR of individual branch near the origin. Specifically, let $\beta$ denote the SNR of the branch, $f(\beta)$ denote the PDF of $\beta$. If
\begin{align}
\lim_{\beta\rightarrow 0}f_\beta(\beta)=a \beta^t +O (\beta^{t+\varepsilon}),
\end{align}
where $O(x)$ is the big O notation, then the high SNR outage probability can be accurate characterized by the parameters $a$ and $t$.

Therefore, the main task is to obtain the behavior of the PDF of $|g|^2$ and $|h_{2,n}|^2 |h_{1,n}|^2$ near the origin. Since $|g|^2$ follows the exponential distribution, its behavior near the origin can be easily obtained. Hence, we focus on $|h_{2,n}|^2 |h_{1,n}|^2$ in the following.

%
Capitalizing on the results of \cite{PDF}, the PDF of random variable $\beta=d_1^{\alpha_1}d_2^{\alpha_2}|h_{2}|^2 |h_{1}|^2$ can be derived as
\begin{align}\label{fbeta}
&f_\beta(\beta) =\notag\\
&2(K_1+1)(K_2+1) e^{-(K_1+K_2)} \sum_{n,p=0}^{\infty}\left(\frac{K_1^{\frac{n}{2}} K_2^{\frac{p}{2}}}{n!p!}\right)^2\times \\
&\left(\sqrt{(K_1+1)(K_2+1)\beta}\right)^{n+p} K_{n-p}\left(2\sqrt{(K_1+1)(K_2+1)\beta}\right).\notag
\end{align}
With the following relationships of bessel-K function \cite[Eq. 10.30.2]{NIST}
\begin{align}
&\lim_{z\rightarrow0}K_{\nu}(z)\approx \frac{1}{2}\Gamma(\nu)(\frac{1}{2}z)^{-\nu}, \quad \Re\,\nu>0,\\
&K_{\nu}(z)=K_{-\nu}(z)\label{sym},
\end{align}
 {where $\Re\,\nu$ represents the real part of $\nu$}. We find it is convenient to consider three separate cases depending on the relationship of $n$ and $p$, namely,
\begin{align}
\lim_{\beta\rightarrow 0}f_\beta(\beta)=P_{n>p}+ {P_{n<p}}+P_{n=p}.
\end{align}

1) For $n>p$, we have
\begin{align}
&P_{n>p} \approx  2(K_1+1)(K_2+1) e^{-(K_1+K_2)}\sum_{p=0}^{\infty}\sum_{n=p+1}^{\infty}  \\ &\left(\frac{K_1^{\frac{n}{2}} K_2^{\frac{p}{2}}}{n!p!}\right)^2
\left(\sqrt{(K_1+1)(K_2+1)\beta}\right)^{2p} \frac{1}{2}\Gamma(n-p).\notag
\end{align}
It is easy to see that $P_{n>p}$ is mainly determined by $p=0$, hence we have
\begin{align}
P_{n>p} \approx  2(K_1+1)(K_2+1) e^{-(K_1+K_2)} \sum_{n=1}^{\infty}\frac{K_1^{{n}} }{2n n!}.
\end{align}

2) For $n<p$, exploiting the symmetric property in Eq. (\ref{sym}), we can similarly derive
\begin{align}
P_{n<p}\approx 2(K_1+1)(K_2+1) e^{-(K_1+K_2)} \sum_{p=1}^{\infty}\frac{K_2^{{p}} }{2p p!}.
\end{align}

3) For $n=p$, we observe that $P_{n=p}$ is dominated by the term $n=p=0$, thus we have
\begin{align}
P_{n=p}\approx &2(K_1+1)(K_2+1) e^{-(K_1+K_2)}\times\notag \\
&K_{0}\left(2\sqrt{(K_1+1)(K_2+1)\beta}\right).
\end{align}

Combining the above three parts, we have $t=0$ and
\begin{align}
a = &(K_1+1)(K_2+1) e^{-(K_1+K_2)}\bigg[{\sf Ei}(K_1)+{\sf Ei}(K_2)-2\gamma\notag\\
&-\ln K_1-\ln K_2+2K_{0}\left(2\sqrt{\frac{(K_1+1)(K_2+1)}{d_1^{\alpha_1}d_2^{\alpha_2}\gamma_0}}\right)\bigg].
\end{align}

To this end, invoking \cite[Eq. (12)]{diversity}, the desired results can be obtained after some algebraic manipulations.

\end{appendices}

\nocite{*}
\bibliographystyle{IEEE}

\begin{thebibliography}{10}

\bibitem{chuang1}
 {C. Huang, S. Hu, G.C. Alexandropoulos, A. Zappone, C. Yuen, R. Zhang, M.D. Renzo, and M. Debbah, ``Holographic MIMO surfaces for 6G wireless networks: Opportunities, challenges, and trends'', [Online]. Available: https://arxiv.org/abs/1911.12296.}


\bibitem{chuang2}
 {C. Huang, R. Mo, and C. Yuen, ``Reconfigurable intelligent surface assisted multiuser MISO systems exploiting deep reinforcement learning'', [Online]. Available:https://arxiv.org/abs/2002.10072.}

\bibitem{Q.Wu1}
Q. Wu and R. Zhang, ``Intelligent reflecting surface enhanced wireless network via joint active and passive beamforming,'' {\em IEEE Trans. Wireless
Commun.}, vol. 18, no. 11, pp. 5394C-5409, Nov. 2019.

\bibitem{Q.Wu2}
Q. Wu and R. Zhang, ``Beamforming optimization for wireless network aided by intelligent reflecting surface with discrete phase shifts,'' accepted to appear in {\em IEEE Trans. Commun.}, 2019.

\bibitem{Huang}
C. Huang, A. Zappone, G. C. Alexandropoulos, M. Debbah, and C. Yuen, ``Reconfigurable intelligent surfaces for energy efficiency in wireless communication,'' {\em IEEE Trans. Wireless Commun.}, vol. 18, no. 8, pp. 4157-C4170, Aug. 2019.


\bibitem{jinshi}
Y. Han, W. Tang, S. Jin, C. Wen, and X. Ma, ``Large intelligent surface-sssisted wireless communication exploiting statistical CSI,'' {\em IEEE Trans. Veh. Technol.}, vol. 68, no. 8, pp. 8238--8242, Aug. 2019.


\bibitem{shuowen}
S. Zhang and R. Zhang, ``Capacity characterization for intelligent
reflecting surface aided MIMO communication,'' \emph{IEEE J. Sel. Areas Commun.}, to appear. [Online]. Available: https://arxiv.org/abs/1910.01573.

\bibitem{X.Hu}
X. Hu, C. Zhong, Y. Zhu, X. Chen, and Z. Zhang, ``Programmable metasurface based multicast systems: Design and analysis,'' accepted to appear in {\em IEEE J. Selected Areas Commun.}, 2020.

\bibitem{zhou}
G. Zhou, C. Pan, H. Ren, K. Wang, and A. Nallanathan, ``Intelligent reflecting surface sided multigroup multicast MISO communication systems,'' [Online]. Available: https://arxiv.org/abs/1909.04606.


\bibitem{cui1}
Y. Jia, C. Ye, and Y. Cui, ``Analysis and optimization of an intelligent reflecting surface-assisted system with interference,'' [Online]. Available: https://arxiv.org/pdf/2002.00168.

\bibitem{cui2}
C. Guo, Y. Cui, F. Yang, and L. Ding, ``Outage probability analysis and minimization in intelligent reflecting surface-assisted MISO systems,'' \emph{IEEE Commun. Lett.}, Early Access.








\bibitem{SISO}
D. Kudathanthirige, D. Gunasinghe, and G. Amarasuriya, ``Performance analysis of intelligent reflective surfaces for wireless communication,'' [Online]. Available: https://arxiv.org/abs/2002.05603.


\bibitem{general}
 {I. Trigui, W. Ajib, and W. Zhu, ``A comprehensive study of reconfigurable intelligent surfaces in generalized fading,'' [Online]. Available: https://arxiv.org/abs/2004.02922.}

\bibitem{J.Gao}
J. Gao, C. Zhong, X. Chen, H. Lin, and Z. Zhang, ``Unsupervised learning for passive beamforming,'' {\em IEEE Wireless Commun. Lett.}, accepted to appear.


\bibitem{Handbook}
M. Abramowitz, and I. A. Stegun, {\em Handbook of mathematical functions}. New York: Dover Publication Inc., 1974.

\bibitem{NIST}
F.~W. Olver, D.~W. Lozier, R.~F. Boisvert, and C.~W. Clark, \emph{{NIST}
  Handbook of Mathematical Functions}.
  New York: Cambridge University Press, 2010.

\bibitem{diversity}
Z. Wang and G. B. Giannakis, ``A simple and general parameterization quantifying performance in fading channels,'' {\em IEEE Trans. Commun.}, vol. 51, no. 8, pp. 1389--1398, Aug. 2003.

\bibitem{PDF}
N. O'Donoughue and J. M. F. Moura, ``On the product of independent complex Gaussians,'' {\em IEEE Trans. Signal Process.}, vol. 60, no. 3, pp. 1050--1063, March 2012.



\end{thebibliography}
\begin{footnotesize}

\end{footnotesize}

\end{document}